\documentclass{trbunofficial}

\usepackage{PRIMEarxiv}
\usepackage[utf8]{inputenc}
\usepackage[T1]{fontenc}
\usepackage{graphicx}
\usepackage{tabularx}
\usepackage{wrapfig}
\usepackage{amsmath,amssymb,amsfonts}
\usepackage{url}
\usepackage{fancyhdr}

\usepackage[colorlinks=true,linkcolor=blue,citecolor=blue,urlcolor=blue]{hyperref}

\AuthorHeaders{Wang, Shahriar, and Zhou}
\title{
Analytical Framework for Evaluating Traffic Capacity Impacts of Electric Vehicles’ Regenerative Braking Dynamics
}

\author{
    Yuhang Wang,
    Md.\ Zidan Shahriar,
    Hao Zhou\thanks{Corresponding author: \texttt{haozhou1@usf.edu}} \\[6pt]
    \small
    Department of Civil and Environmental Engineering, University of South Florida, Tampa, FL 33620, USA
}

\begin{document}
\maketitle

\begin{abstract}
Regenerative braking (RB) significantly influences electric vehicle (EV) car-following (CF) dynamics, yet traditional traffic-flow models rarely capture these effects. We introduce a comprehensive empirical dataset comprising 197.5 hours of driving data from 25 drivers across eight EV models to systematically investigate regen-induced CF behaviors. Two primary CF patterns emerge: (i) steady-state scenarios where EVs use regenerative braking and subsequently re-accelerate to equilibrium speeds with larger spacing, and (ii) dynamic scenarios involving lead oscillations, characterized by a distinctive three-phase CF process—regenerative deceleration, transitional plateau, and rapid re-acceleration. The paper's main contribution is the development of an analytical framework that models these EV-specific CF behaviors and quantifies their impacts on traffic capacity. We derive closed-form expressions for the established $\eta$ function from the literature, explicitly quantifying EV driving deviations from stable CF defined by Newell's CF model and assessing their implications for roadway capacity. Validation against empirical data and simulation confirm the model's accuracy ($R^2=0.96$) in replicating real-world $\eta$ trajectories. Sensitivity analyses demonstrate that increased RB intensity, prolonged transitions, and shorter reaction delays significantly raise values and cumulative capacity losses. These findings highlight a clear trade-off between enhanced energy recovery through RB and reduced traffic efficiency, providing critical insights for EV-aware traffic modeling, control strategies, and transportation policy.
\end{abstract}

\keywords{Electric Vehicle \and Regenerative Brake \and Car Following \and Traffic Flow Impacts}

\section{Introduction}


Car‐following (CF) models bridge microscopic driving behaviour and macroscopic traffic laws such as flow, density, and capacity, providing a unified language for theory and practice \cite{laval2010mechanism, chen2012behavioral, zhong2024understanding}. While classical models reproduce the growth of stop–go waves in ICEV platoons, their calibration implicitly assumes symmetric acceleration–deceleration dynamics. Over the past decades, an extensive family of CF models has been developed to capture driver heterogeneity, stochastic effects, and the advent of vehicle automation. Representative examples include the Improved Intelligent Driver Model (IIDM), which adjusts IDM’s equilibrium gap to better match measured acceleration behaviour \cite{kesting2010enhanced}; the asymmetric-behaviour (AB) model, which attributes stop–go growth to distinct acceleration and deceleration responses \cite{CHEN2012744}; two-regime stochastic formulations that switch between free-flow and congested states under Brownian excitation \cite{XU2020210}; and deep-learning surrogates that infer nonlinear interactions directly from trajectory data \cite{xie2020deepcf}. Although these models excel at reproducing detailed dynamics, their mathematical complexity often clouds the link to macroscopic quantities. By contrast, Newell’s lag–spacing rule retains analytic clarity: a follower trajectory is generated from the leader’s by a fixed time delay $\tau$ and spatial shift $\delta$ \cite{newell2002simplified}. This straight-line mapping embeds the triangular fundamental diagram, making Newell’s model a convenient micro–macro bridge for capacity estimation, shock-wave analysis, and, more recently, EV-specific traffic assessment \cite{ZHOU2022103801, fan2013data}. Its closed-form metrics—most notably the space–time filling ratio $\eta(t)$ and the cumulative capacity loss $\mathcal{L}$—are exploited in this study to translate EV-specific kinematics, particularly fixed-rate regenerative braking and motor-driven acceleration, into network-level performance measures.

EVs are reshaping this landscape. EVs already account for more than 10\,\% of U.S.\ new-vehicle sales and are poised for continued growth \cite{Porfirenko2025,IEA2025}. Key to their distinct longitudinal motion is \emph{driver-selectable regenerative braking}: a quasi-constant negative acceleration engaged immediately upon throttle release, whose magnitude increases monotonically with the chosen regen level \cite{yang2024regenerative}. Laboratory and on-road tests report steady decelerations on the order of  0.1–0.30 g, prompting drivers to lift earlier and maintain larger headways—behaviour rarely observed in internal-combustion cohorts \cite{Oleksowicz01052013}. However, most microscopic CF and macroscopic capacity models remain calibrated on ICEV data or ignore EV's impact and therefore risk misestimate stability and throughput in EV-rich streams. 


This study closes that gap. Leveraging \textbf{197.5 h} of high-resolution trajectories from eight production EV models—Tesla Model 3/X/Y, Hyundai Ioniq 5, Kia Niro/EV6, Ford Mach-E, Chevrolet Bolt—plus complementary community runs, we: (i) quantify the fixed deceleration and acceleration plateaus, as well as the finite‐slope transition segment that connects them; (ii) embed these empirically derived dynamics into an EV-specific extension of Newell’s lag–spacing rule, deriving closed-form expressions for the space–time filling ratio $\eta(t)$ and the cumulative capacity loss $\mathcal{L}$ under both constant-speed and sinusoidally decelerating leaders; (iii) validate the analytic formulae against controlled experiments and naturalistic data, achieving an \textbf{$R^2=0.96$} fit to real EV trajectories; (iv) A sensitivity study demonstrates that selecting a stronger regenerative-braking setting, responding more rapidly to the leader’s speed change, and keeping a longer cruising interval between braking and acceleration all enlarge headway fluctuations and increase the resulting capacity loss. These insights offer concrete guidance for choosing default regen levels and for designing cooperative, corridor-wide control strategies.

The resulting framework furnishes a physics-grounded bridge between driver-selected regen settings and macroscopic traffic efficiency, laying a tractable foundation for EV-aware simulation, infrastructure assessment, and cooperative control design.

\section{EV Kinematic Dataset}

\begin{figure}[!ht]
  \centering
  \includegraphics[width=0.9\textwidth]{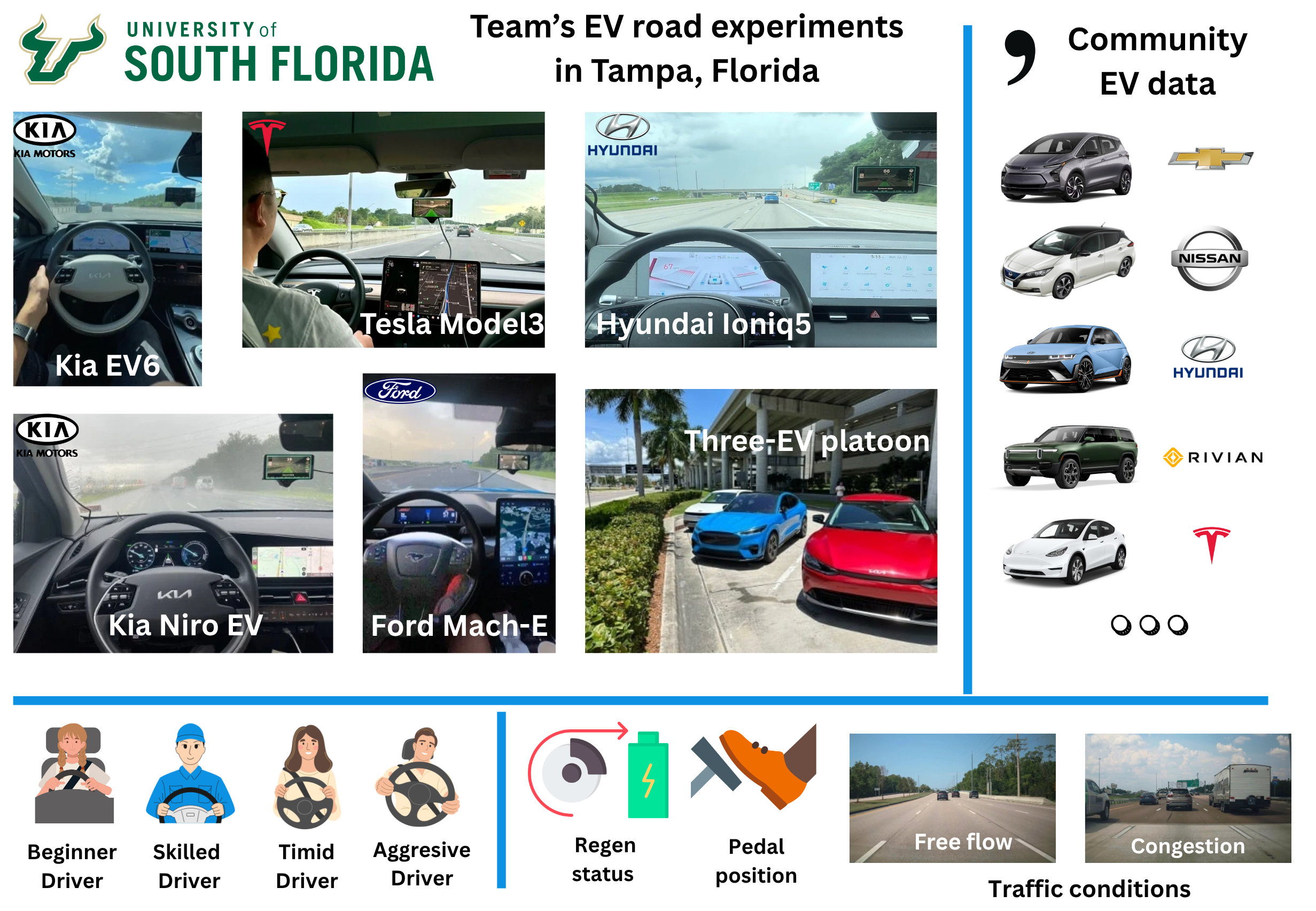}
  \caption{EV Dataset: 150 h of naturalistic driving collected since 2024 by four volunteers on six EV models. Comma-community dataset: daily logs from 21 drivers, adds a further 20 h and expands coverage to 9 EV models. Together the two datasets provide 197.5 h of high-resolution data across $>$ 10000 miles, spanning novice ($<$ 1 month) to highly experienced ($>$ 10 years) drivers, timid and aggressive styles, and both free-flow and congested conditions.}
  \label{fig:Dataset}
\end{figure}

Considering that current CF datasets primarily capture vehicle trajectories using roadside cameras\cite{gloudemans202324}, drones\cite{XU2021116356}, and other tools, they lack detailed control and perception data of both the leading and following vehicles over long periods of time, making it difficult to capture the completeness and accuracy of the entire car-following process. Other studies use closed-source test data\cite{LI202167}, which often focuses on a single vehicle type, scenario, model, and driver. Most tests are based on gasoline vehicles, making them inadequate for electric vehicle following. Furthermore, the lack of comprehensive access to these datasets limits research due to closed-source data. To address these gaps, we rented various electric vehicle models for daily driving at Tampa International Airport starting in June 2024, including the Tesla Model 3, X, and Y; the Kia Niro and Kia EV 6; the Hyundai Ioniq; and the Ford Mach-E, totaling 144.9 hours of data. To enrich the data on drivers and electric vehicle models, we publicly accessed 52.6 hours of daily driving data from around the world from the Comma Community. We expanded the data to nine electric vehicle models and 25 drivers.

We recorded our own vehicle's driving data using a professional vehicle logging tool and installed a dash camera to capture driving data from the vehicle ahead. Through comprehensive data processing, we reverse-engineered the logs to obtain high-frequency and refined speed, acceleration, pedal (accelerator) position, brake status/position, and other information. We also used the dash cam to obtain accurate motion data (speed, distance) from the preceding vehicle using the advanced end-to-end model OpenPilot. As shown in Fig.~\ref{fig:Dataset}, given that a single driver or a single brand of electric vehicle cannot capture the universal characteristics of drivers, our dataset covers driving data from a large number of drivers across various electric vehicle models. We even categorize drivers by style, ranging from those who have just obtained their license (less than a week) to those with over 10 years of driving experience; from those new to electric vehicles to those who are experienced; and even includes Timid. driver and aggressive driver. This dataset has been made public on Github and can be accessed through the link \url{https://github.com/OpenLKA/EV_Dataset}.

\section{Characteristics of EV Kinematics}

EV kinematics diverge from those of internal-combustion engine vehicles(ICEVs) most visibly in the acceleration and deceleration phases. Higher, instantaneously available motor torque and the absence of gear-shift delays give EVs greater peak acceleration and shorter response lags, whereas ICEVs rely on throttle-induced torque build-up. Conversely, deceleration in EVs is dominated by driver-selectable regenerative braking, which supplies a near-constant negative acceleration unavailable in friction-only ICEVs. In what follows, we incorporate these two asymmetries into the car-following framework and quantify EV-specific parameters using dynamometer measurements and naturalistic trajectory data.

\subsection{EV's Regen Deceleration}

\begin{wrapfigure}{r}{0.3\textwidth}
  \centering
  \includegraphics[width=0.25\textwidth]{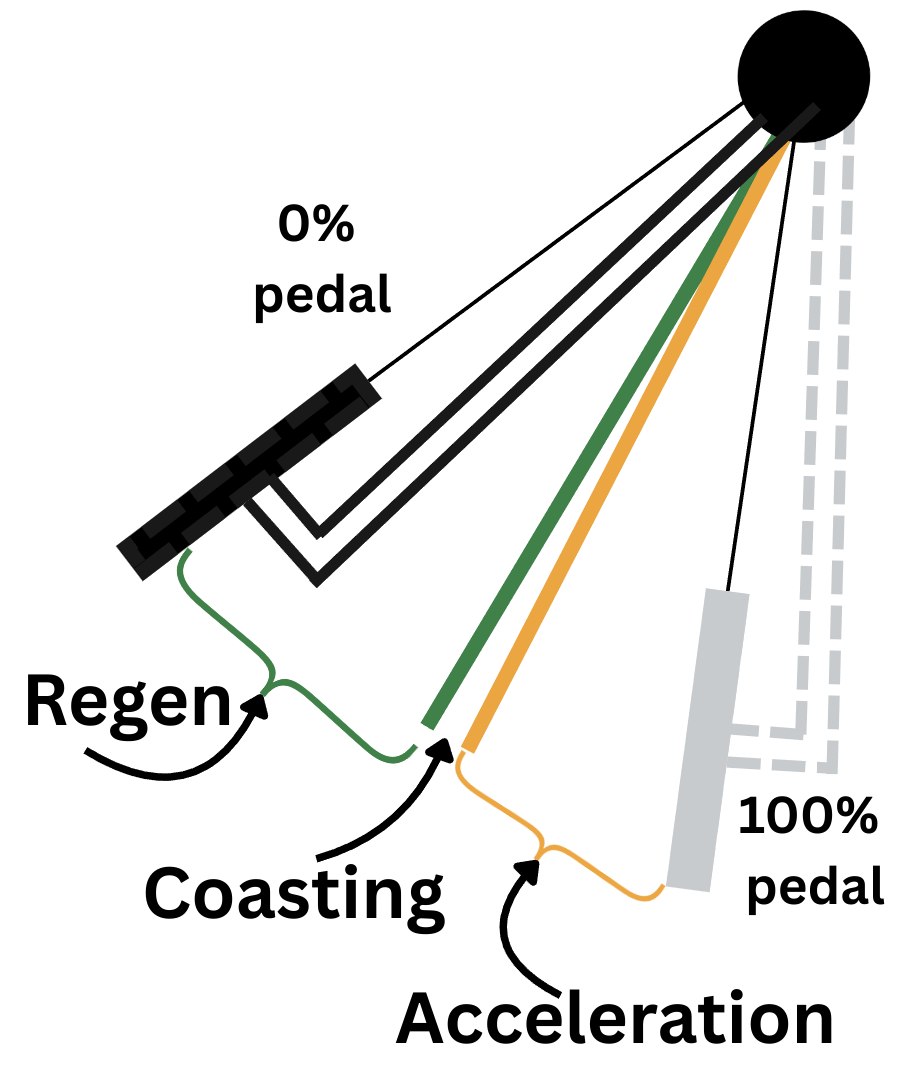}
  \caption{Acceleration Range of Tesla in one-pedal mode}
  \label{fig:regen_range}
\end{wrapfigure}

RB enables EVs to convert wheel torque into electrical energy almost instantaneously by driving the traction motor in generator mode; modern model-predictive, fuzzy, and game-theoretic controllers maximise this recuperation while preserving vehicle stability. The resulting high, context-blind regeneration torque suppresses natural coasting and imposes a quasi-constant deceleration profile-typically up to $\approx$ 0.3 g-that reshapes longitudinal kinematics and amplifies car-following variability. Reflecting these characteristics, many mass-market EVs now default to a one-pedal driving (OPD) scheme in which the accelerator alone governs both propulsion and RB: partial or full pedal release at cruising speed immediately triggers regenerative torque, while the stroke range itself divides into three contiguous bands-acceleration, a narrow coasting sliver, and deceleration-whose boundaries shift with vehicle speed (Fig. \ref{fig:regen_range}).

\begin{wrapfigure}{l!h}{0.5\textwidth}
  \centering
  \includegraphics[width=0.48\textwidth]{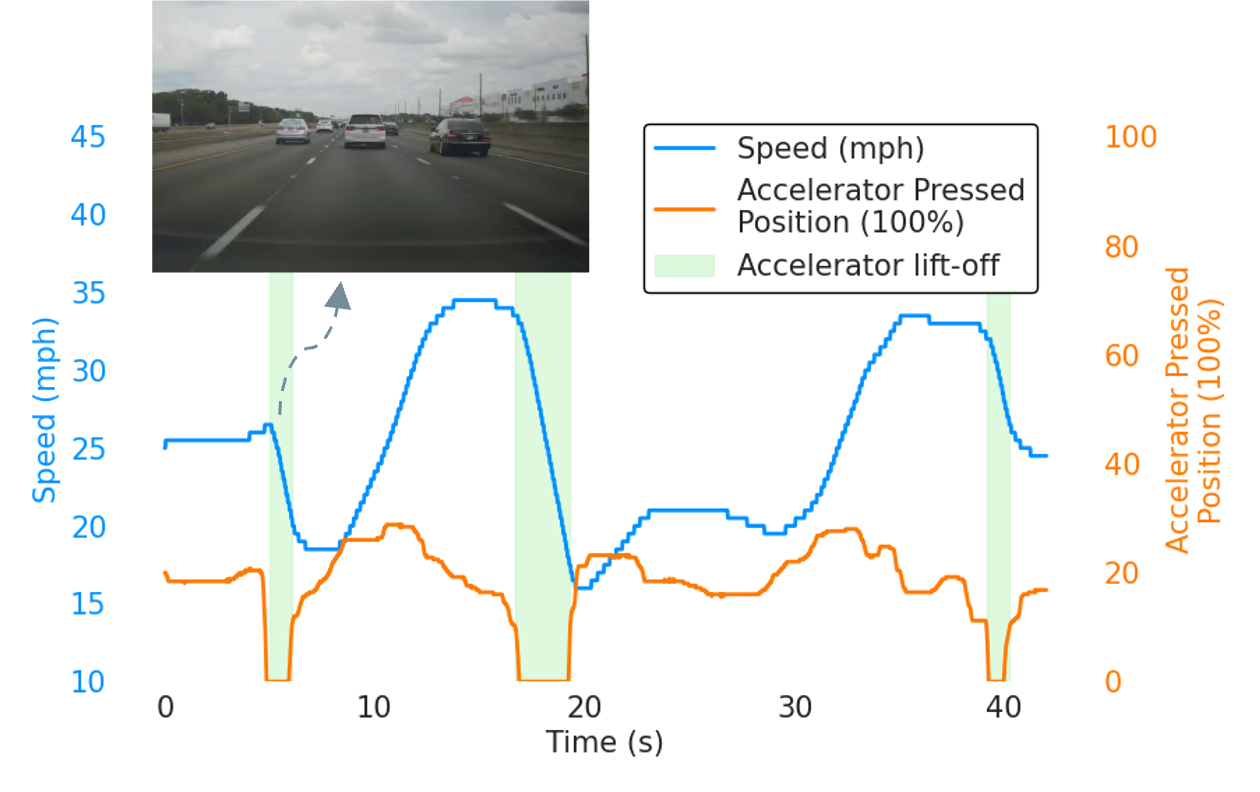}
  \caption{Regen engages abrupt and strong deceleration upon pedal releases (a 2025 Tesla Model 3)}
  \label{fig:abrupt_decel}
\end{wrapfigure}

EV shows a high pedal sensitivity. A distinctive consequence of one-pedal driving is the extremely steep mapping between accelerator-pedal stroke and longitudinal force. Laboratory roller-dyno tests and on-road measurements show that, for many late-model EVs, \emph{less than one-fifth} of the total pedal travel covers the entire span from peak propulsion to peak RB \cite{belt2020tesla}.  Figure~\ref{fig:abrupt_decel} illustrates a representative highway instance: when the driver of a Tesla Model~3 releases the pedal from a modest 15\,\% position while cruising at $\approx$\,28\,mph in moderately congested flow, the vehicle sheds nearly 10\,mph in just one second, corresponding to an average deceleration of $\sim$\,0.15\,g.  Because such large kinematic swings arise from minute pedal motions, unintentional micro-oscillations of the driver’s ankle-or simply over-eager throttle corrections-translate into pronounced acceleration “overshoots" and deceleration “undershoots."  These fluctuations increase the standard deviation of speed within a platoon, intensify the growth rate of stop-and-go waves, and ultimately reduce lane throughput, a mechanism already recognised in ICEV traffic as a root cause of oscillatory instabilities \cite {laval2010mechanism}.  The situation is further complicated by the absence of a design consensus across manufacturers: while most OEMs multiplex propulsion and RB on the right-hand pedal, others (e.g., Chevrolet Bolt, Honda~e) offer steering-wheel paddles to modulate RB levels, introducing yet another layer of driver-machine adaptation and potentially widening the spectrum of behavioural responses \cite{honda2024regenpaddle}.  Although longitudinal-learning studies suggest that drivers gradually tame their pedal inputs after several months of exposure, our empirical data indicates that abrupt lift-offs remain frequent even among experienced owners, underscoring the persistence of high pedal sensitivity as a traffic-relevant trait.

\begin{figure}[!h]
  \centering
  \includegraphics[width=0.8\textwidth]{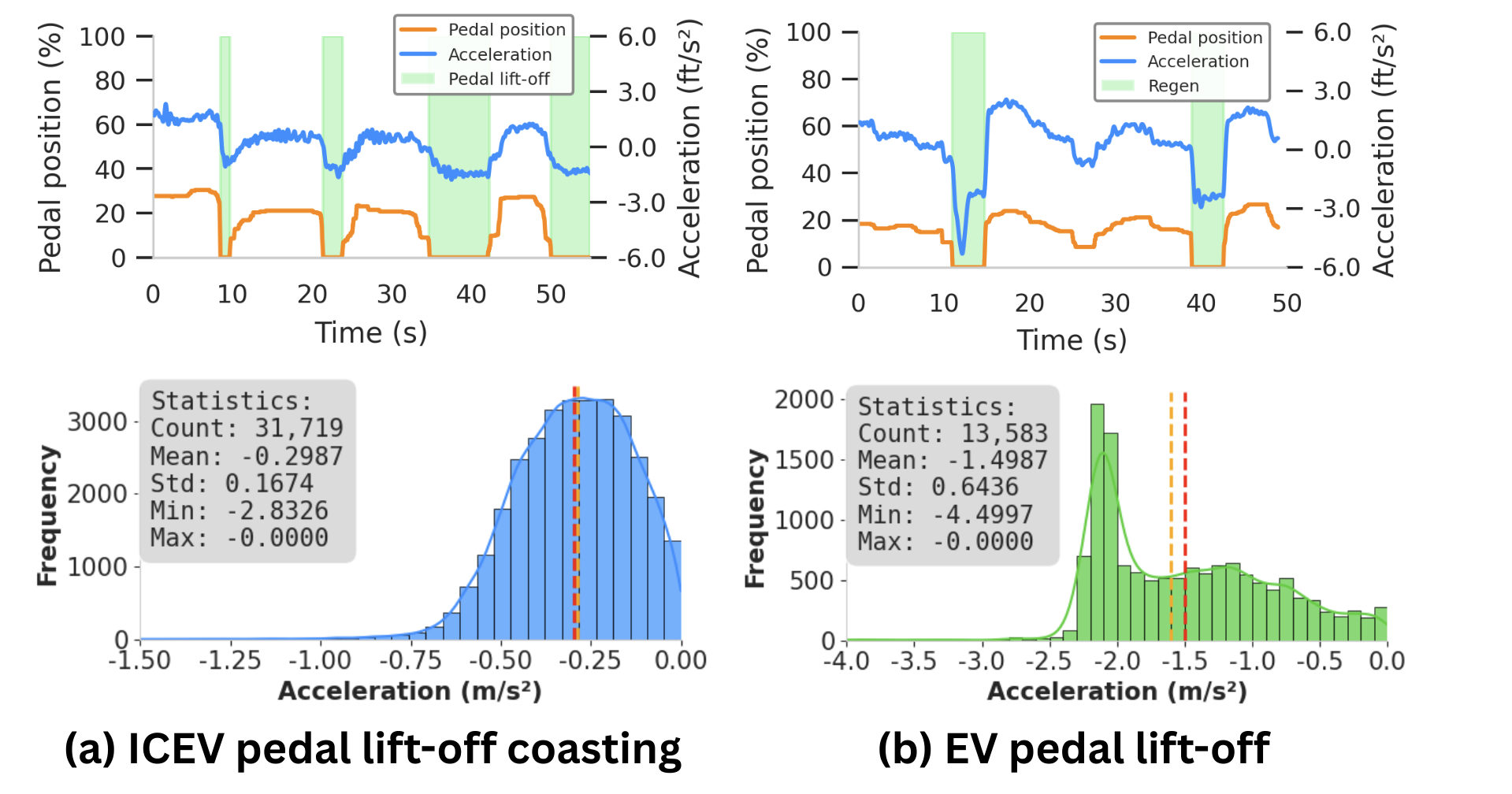} 
  \caption{Comparison of EV(Tesla Model3) and ICEV(Honda civic) at accelerator pedal lift-of, ICEV shows a coasting profile while EV reveals high constant regen deceleration.}
  \label{fig:ev_icev_brake_decel}
\end{figure}

The inertia-driven coasting buffer commonly seen in ICEV, disappear in EV's driving. In an ICEV, releasing the accelerator cuts fuel injection, but the driveline remains engaged; the car therefore coasts under aerodynamic drag and rolling resistance alone, losing speed at a gentle 0.02-0.05 g.  This passive buffer absorbs small speed perturbations, allowing a following human driver to smooth out micro-fluctuations and damp upstream disturbances.  Under one-pedal RB, by contrast, the power-train inverts almost instantaneously to generator mode: regenerative torque peaks within 200-300 ms of pedal lift-off, enforcing a quasi-constant deceleration up to $\sim$\,0.3 g \cite {LEE2023128745}.  Figures~\ref{fig:abrupt_decel} and~\ref{fig:ev_icev_brake_decel} juxtapose the resulting speed traces for an EV and a class-matched ICEV subjected to identical throttle releases; the EV trace plunges almost vertically, whereas the ICEV trace resembles a shallow ramp.  The near-absence of a coasting phase in EVs removes the natural “buffer time" that typically shields a driver from small lead-vehicle speed changes.  As a result, EV followers must choose between accepting larger static headways or engaging in more frequent micro-braking, either of which alters the equilibrium spacing-speed relation fundamental to microscopic car-following models.  From a macroscopic standpoint, the vanishing coasting buffer can retard platoon discharge at signalised intersections (when multiple EVs lift simultaneously) yet also damp the acceleration of backward-propagating shockwaves in mild congestion, producing a nuanced, density-dependent impact on overall capacity-an effect quantified in Section~\ref{sec:modeling}.

\subsection{EV's constant acceleration and deceleration}

\begin{figure}[htbp]
  \centering
  \includegraphics[width=0.9\textwidth]{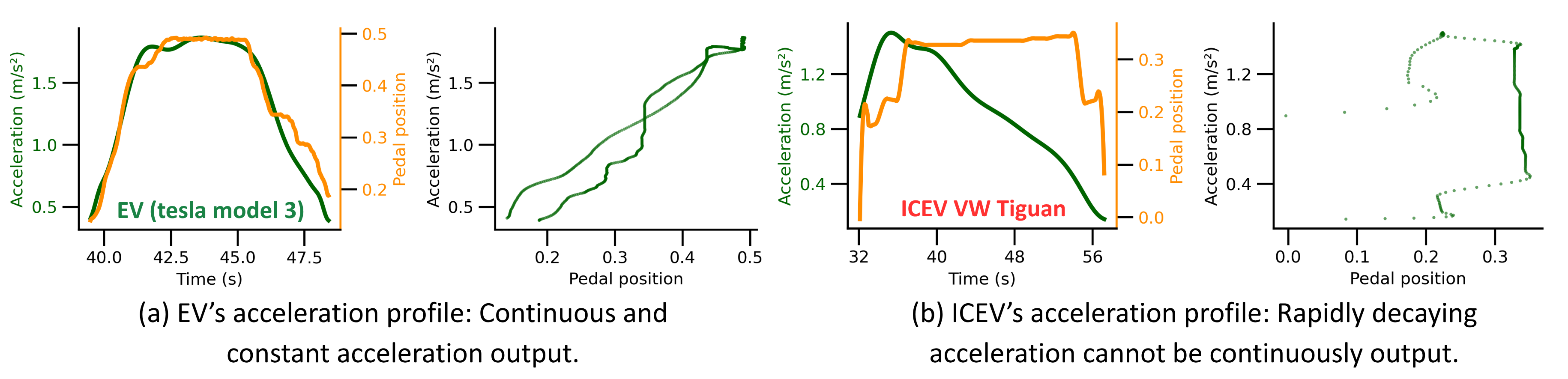} 
  \caption{EV's continues constant acceleration and ICEV's quick acceleration decay}
  \label{fig:acceleration_profile_comparison}
\end{figure}

Compared with ICEVs, EVs sustain a nearly constant peak-level acceleration over a much broader speed range because the traction motor can deliver maximum torque almost instantaneously and maintain high power without gear-shift interruptions \cite{meyn2015kinematic}.  Figure~\ref{fig:acceleration_profile_comparison} illustrates that, under comparable throttle inputs, an EV swiftly attains its peak longitudinal acceleration and holds this level for several seconds, whereas the ICEV’s acceleration decays rapidly once engine torque and gear ratios move away from their optimum points.  To capture this salient difference, we approximate the EV’s launch phase as a period of uniform acceleration in the subsequent model development; this assumption preserves physical interpretability while greatly simplifying the analytical formulation.

Not only do EVs exhibit a fixed acceleration pattern during acceleration, but when analyzing their RB deceleration profiles, we found that they also exhibit a fixed deceleration pattern. As shown in Fig.~\ref{fig:acceleration_profile_comparison}, for four different EV models, with the accelerator nearly lifted, RB provides fixed deceleration values ranging from 1 $m/s^2$ to 4 $m/s^2$, depending on the user's preset RB strength.

\begin{figure}[!h]
  \centering
  \includegraphics[width=0.8\textwidth]{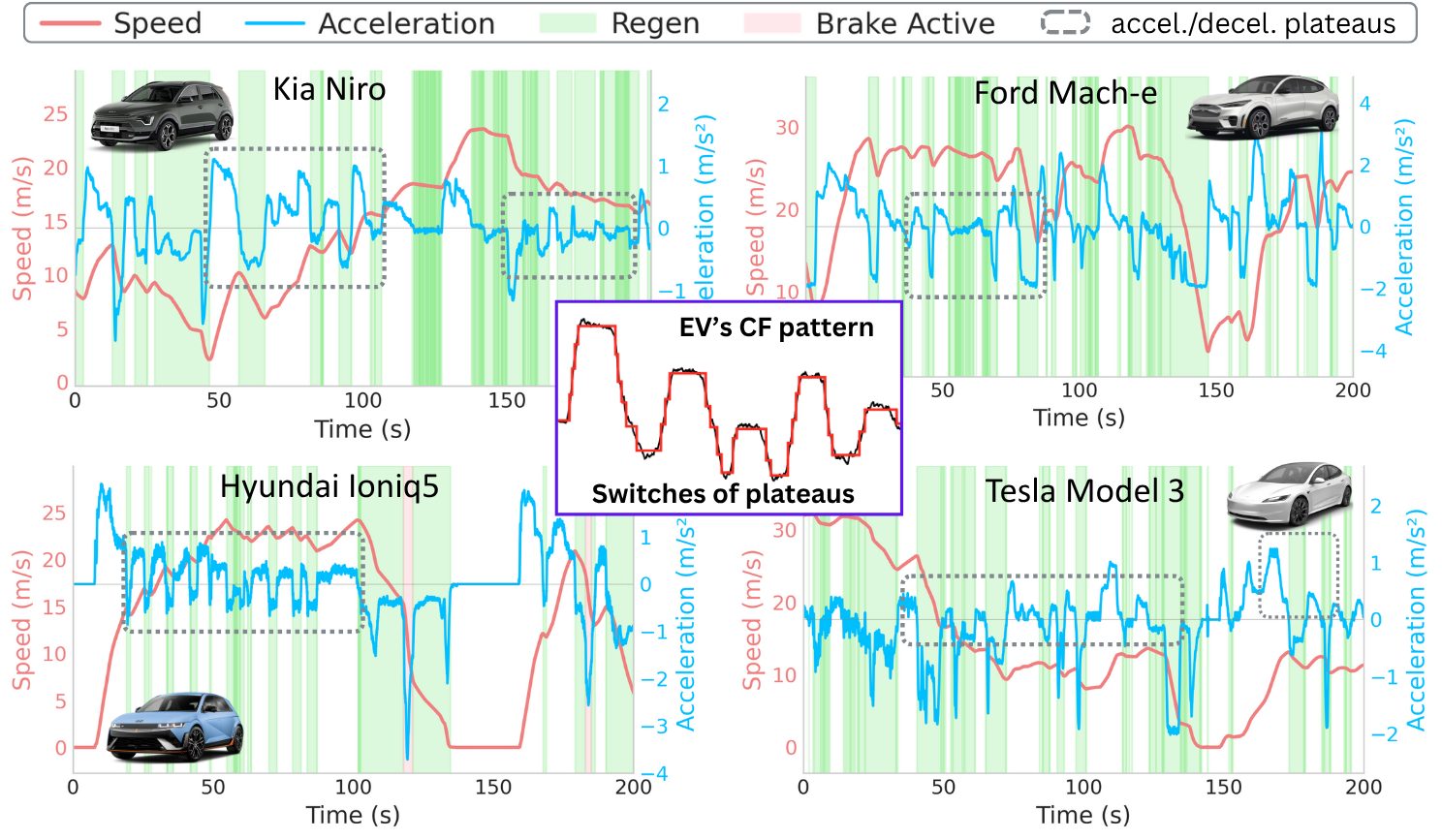} 
  \caption{Typical EV kinematics in daily driving, EV shows an constant acceleration and deceleration value and a transition segment between acceleration changes.}
  \label{fig:kinematic}
\end{figure}

\subsection{Transition between deceleration and acceleration}

\begin{figure}[!h]
  \centering
  \includegraphics[width=1.\textwidth]{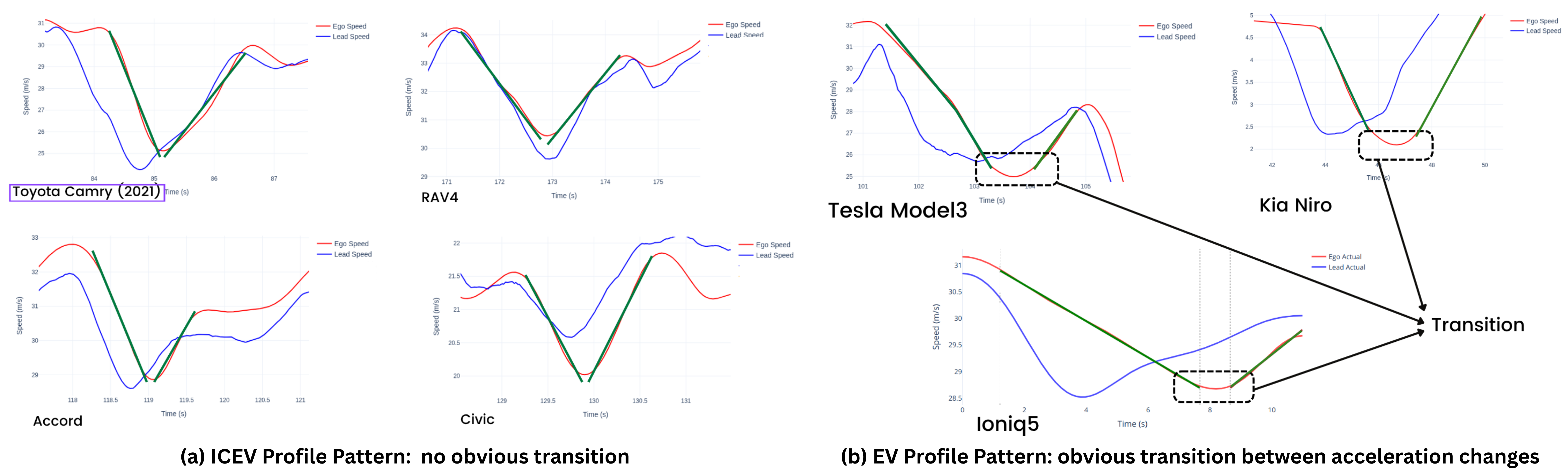} 
  \caption{In the transition section between acceleration and deceleration, EV shows an obvious flat transition pattern in (b) while ICEV don't have similar transition in (a).}
  \label{fig:transition}
\end{figure}

Most EV manufacturers consider safety and comfort when designing their regenerative deceleration and acceleration systems. Therefore, they implement a jerk limit to prevent excessive jerk during acceleration and deceleration. As shown in Fig.~\ref{fig:transition}, the transition between acceleration and deceleration exhibits a brief, constant speed phase. This transition ensures smooth driving. ICEVs, on the other hand, lack a distinct constant acceleration or deceleration, and therefore lack a transition.

In summary, the empirical data reveals the following kinematic characteristics of EVs: 1) EVs exhibit high pedal sensitivity, enabling them to generate significant acceleration immediately upon release of the accelerator, thus eliminating the coasting characteristic of ICEVs, which often results in overshooting and undershooting. 2) EVs maintain a constant acceleration throughout acceleration and deceleration. We will utilize these characteristics to model the CF of EVs in the next section. This allows us to explore how these characteristics affect vehicle following and traffic flow.

\section{EV CF Modeling}
\label{sec:modeling}

In this section, we first define two basic EV CF models. We then use the deviation ($\eta$) from the Newell's CF model to characterize the impact of these following modes on traffic flow. Finally, we explicitly model $\eta$ and capacity loss using the physical parameters of the electric vehicle, helping us quantify how different factors affect traffic flow.

\subsection{Scenario Definition: EV CF}

Considering that while the proportion of EVs in traffic is increasing, the majority of vehicles are still ICEVs, a CF model where EVs follow ICEVs or ICEVs follow EVs is generally more universal. Consider modeling the following two scenarios:

\begin{figure}[htbp]
  \centering
  \includegraphics[width=0.9\textwidth]{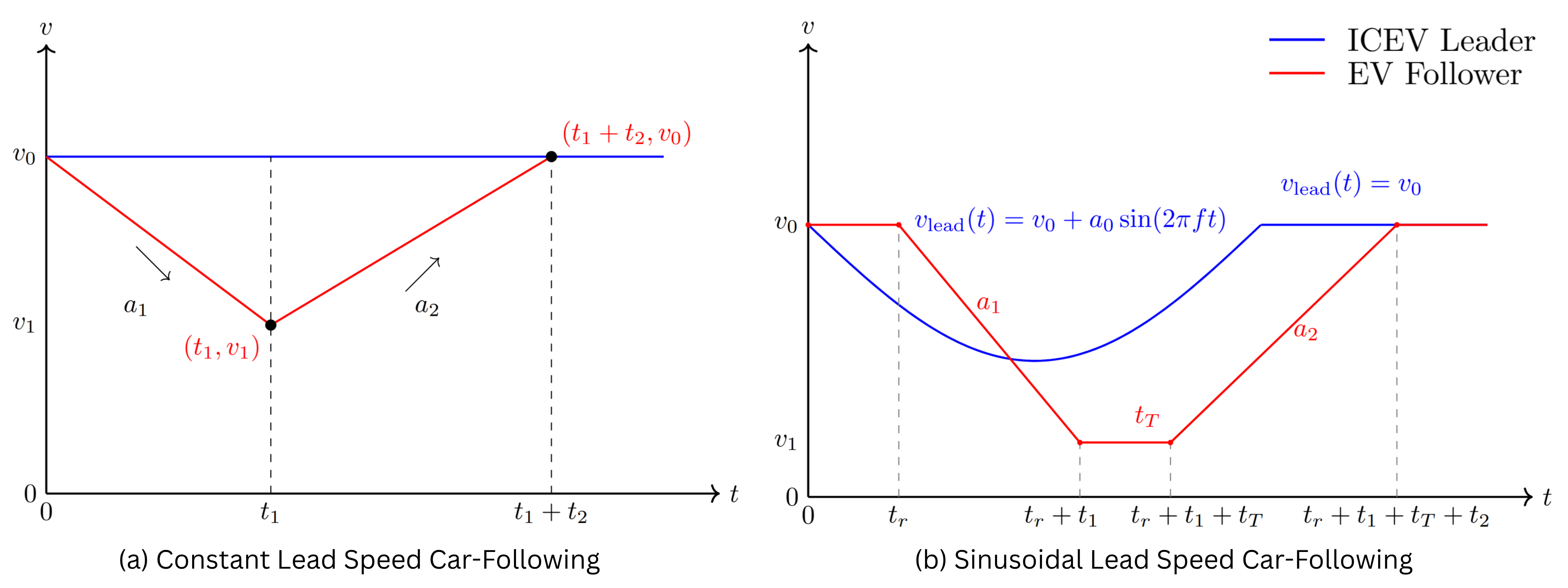} 
  \caption{Speed Profile of 2 CF Scenarios; Scenario 1: Constant-speed leader, regen-brake-recover follower; Scenario 2: Leader brake-recover, regen-brake-recover follower}
  \label{fig:scenarios}
\end{figure}

\textbf{Constant-speed leader, regen-brake-recover follower} This scenario is shown in Fig.~\ref{fig:scenarios} (a). The EV follower (following vehicle) and the ICEV leader (leading vehicle) are both traveling at a initial speed of $v_0$ (typically the road speed limit), with an initial following distance of $s_0$. At time $t=0$, the EV follower activates the regen brake, causing the vehicle to decelerate at an acceleration of $a_1$ for $t_1$. During this time, the following distance increases, and the EV follower then accelerates at an acceleration of $a_2$ until it returns to its original speed of $v_0$ at time $t_2$. Throughout this process, the ICEV leader maintains a constant speed of $v_0$. The speed profile of the leader and the follower can be presented in Eq.~\ref{eq:speed_leader} and Eq.~\ref{eq:speed_follower}.

\begin{linenomath}
  \begin{equation}
  \label{eq:speed_leader}
    v_{\mathrm{lead}}(t)=v_0,\qquad t\ge 0 .
  \end{equation}
\end{linenomath}

\begin{linenomath}
  \begin{equation}
  \label{eq:speed_follower}
    v_{\mathrm{follower}}(t)=
    \begin{cases}
      v_0 + a_1\,t, & 0\le t<t_1,\\[4pt]
      v_1 + a_2\,(t-t_1), & t_1\le t<t_2,\\[4pt]
      v_0, & t\ge t_2,
    \end{cases}
  \end{equation}
\end{linenomath}

Thus, we can get the trajectory of the ICEV leader and EV follower in the following equations (Eq.\ref{eq:traj_leader} and Eq. \ref{eq:traj_follower}):

\begin{linenomath}
  \begin{equation}
  \label{eq:traj_leader}
    x_{\mathrm{lead}}(t)=v_0  \cdot t + s_0,\qquad t\ge 0 .
  \end{equation}
\end{linenomath}

\begin{linenomath}
  \begin{equation}
  \label{eq:traj_follower}
    x_{\mathrm{follower}}(t)=
    \begin{cases}
      v_0\,t+\dfrac12 a_1 t^2, & 0\le t \le t_1,\\[2pt]
      x_{follower}(t_1)+v_1(t-t_1)+\dfrac12 a_2(t-t_1)^2, & t_1 < t<t_2.
    \end{cases}
  \end{equation}
\end{linenomath}

This constant-speed-leader and regen-brake-recover-follower configuration aptly captures two common situations: (i) a cautious EV driver who deliberately taps regenerative braking to enlarge the headway to the vehicle in front, and (ii) the brief deceleration pulses produced by small pedal oscillations when the car is driven in one-pedal mode. Furthermore, when $a_1 > 0$ and $a_2 < 0$, the EV follower exhibits a speed profile that accelerates first and then decelerates, while the ICEV maintains a constant speed. In this case, this model can be used to characterize the situation where the EV follower catches up with the ICEV leader.

\textbf{Leader brake-recover, regen-brake-recover follower} While the previous scenario effectively models the following behavior of a leading vehicle traveling at a constant speed, the ego vehicle generally does not decelerate when the leading vehicle exhibits no speed fluctuations. Therefore, in this section, we simulate the ICEV leader's initial deceleration and subsequent recovery to its initial speed, thereby dynamically modeling the following vehicle's behavior. As shown in Fig.~\ref{fig:scenarios} (b), we first use the commonly used sinusoidal curve (ranney2009measuring, wang2023deep) to construct the leading vehicle's speed profile, as shown in Eq.~\ref{eq:sinusoidal_lead_speed}. Its accuracy and simplicity in modeling gasoline vehicle speed profiles have made it widely used. In this scenario, the ICEV follower's initial speed is $v_0$. The ICEV follower then undergoes a half-cycle of oscillations, from deceleration to acceleration, and finally to recovery to its initial speed, with an amplitude of $a_0$ and a frequency of $2f$. Its trajectory equation corresponds to Eq.~\ref{eq:sinusoidal_lead_traj}. For the EV follower, its velocity equation is Eq.~\ref{eq:sinusoidal_follower_speed}. 

\begin{linenomath}
  \begin{equation}
    \label{eq:sinusoidal_lead_speed}
    v_{\mathrm{lead}}(t)=
    \begin{cases}
      v_0 + a_0\,\sin(2\pi f t), & 0 \le t \le \dfrac{1}{2f},\\[4pt]
      v_0, & t > \dfrac{1}{2f}.
    \end{cases}
  \end{equation}
\end{linenomath}

\begin{linenomath}
  \begin{equation}
    \label{eq:sinusoidal_lead_traj}
    x_{\mathrm{lead}}(t)=
    \begin{cases}
      s_0 + v_0 t + \dfrac{a_0}{2\pi f}\bigl[1-\cos(2\pi f t)\bigr], & 0 \le t < \dfrac{1}{2f},\\[6pt]
      s_0 + \dfrac{a_0}{\pi f} + v_0 t, & t \ge \dfrac{1}{2f}.
    \end{cases}
  \end{equation}
\end{linenomath}

The EV also starts with an initial velocity $v_0$ and moves at a constant speed until it detects the preceding vehicle's deceleration at time $t_r$ (reaction time). Given the EV's constant acceleration characteristic, the EV follower will decelerate uniformly at a deceleration of $a_1$ for $t_1$ until its speed drops to $v_1$. The driver will then accelerate the vehicle to return to $v_0$. Considering the relatively smooth transition pattern between deceleration and acceleration in EVs, the EV follower will maintain a constant velocity of $v_1$ for $t_T$ (transition time), followed by a uniform acceleration of $a_2$ until its speed returns to its initial velocity $v_0$. Its trajectory equation corresponds to Eq.~\ref{eq:sinusoidal_follower_traj}. This scenario effectively accounts for the dynamic characteristics of EVs and helps analyze the differences between EV and ICEV following. Next, we will describe the method we used to analyze the impact of the following model on traffic flow.

\begin{linenomath}
  \begin{equation}
    \label{eq:sinusoidal_follower_speed}
    v_{\mathrm{follower}}(t)=
    \begin{cases}
      v_0, & 0 \le t < t_r,\\[4pt]
      v_0 + a_1\,(t-t_r), & t_r \le t < t_r+t_1,\\[4pt]
      v_1, & t_r+t_1 \le t < t_r+t_1+t_T,\\[4pt]
      v_1 + a_2\,(t-t_r-t_1-t_T), & t_r+t_1+t_T \le t < t_r+t_1+t_T+t_2,\\[4pt]
      v_0, & t \ge t_r+t_1+t_T+t_2.
    \end{cases}
  \end{equation}
\end{linenomath}

\begin{linenomath}
  \begin{equation}
    \label{eq:sinusoidal_follower_traj}
    x_{\mathrm{follower}}(t)=
    \begin{cases}
      v_0\,t, & 0 \le t < t_r,\\[4pt]
      v_0\,t + \dfrac{a_1}{2}\,(t-t_r)^2, & t_r \le t < t_r+t_1,\\[6pt]
      x_{\mathrm{follower}}(t_r+t_1)+v_1\,(t-t_r-t_1), & t_r+t_1 \le t < t_r+t_1+t_T,\\[6pt]
      x_{\mathrm{follower}}(t^\ast)+v_1\,(t-t^\ast)+\dfrac{a_2}{2}\,(t-t^\ast)^2, & t^\ast \le t < t^\ast+t_2,\\[6pt]
      x_{\mathrm{follower}}(t^\ast+t_2)+v_0\,(t-t^\ast-t_2), & t \ge t^\ast+t_2,
    \end{cases}
  \end{equation}
\end{linenomath}
\[
\text{with}\quad 
t^\ast = t_r+t_1+t_T,\quad
v_1 = v_0 + a_1 t_1.
\]

As a convenience, Table~\ref{tab:notation} summarises all symbols introduced so far together with their physical meanings and units.

\begin{table}[!ht]
  \caption{Notation used in the EV--ICEV car‐following models}
  \label{tab:notation}
  \begin{center}
    \begin{tabularx}{\linewidth}{l X l}
      \textbf{Symbol} & \textbf{Meaning} & \textbf{Units / note} \\\hline
      $v_0$  & Initial (cruise) speed of both vehicles & m\,s$^{-1}$ \\
      $s_0$  & Initial spacing (ICEV $\rightarrow$ EV) & m \\
      $a_1<0$ & EV regenerative deceleration & m\,s$^{-2}$ \\
      $a_2>0$ & EV catch-up acceleration & m\,s$^{-2}$ \\
      $a_0<0$ & Amplitude of leader’s sinusoidal decel & m\,s$^{-2}$ \\
      $f$     & Frequency of leader sine wave ($1/2f$ half-period) & Hz \\
      $t_1$   & Duration of EV decel at $a_1$ & s \\
      $t_2$   & Duration of EV accel at $a_2$ & s \\
      $t_r$   & Driver reaction time (EV) & s \\
      $t_T$   & Coasting / transition time at $v_1$ & s \\
      $t^\ast$& $t^\ast=t_r+t_1+t_T$ (start of accel) & s \\
      $v_1$   & $v_1=v_0+a_1 t_1$ (intermediate speed) & m\,s$^{-1}$ \\
      $\delta$& Newell spatial shift & m \\
      $\tau$  & Newell time lag & s \\
      $\omega$& Lag-spacing slope $\omega=\delta/\tau$ & m\,s$^{-1}$ \\
      $\eta(t)$ & Space-time ratio (Eqs.~\eqref{eq:eta_def}, \eqref{eq:eta_scenario1}) & -- \\
      $\mathcal{L}$ & Cumulative capacity loss (Eq.~\eqref{eq:caploss_int}) & veh·s (lane·m) \\
      $t_A$   & Intersection time on leader trajectory & s \\
      $t_C$   & Observation time on follower trajectory & s \\\hline
    \end{tabularx}
  \end{center}
\end{table}

The symbols in Table~\ref{tab:notation} are used throughout the remainder of the paper without further clarification.

\subsection{Newell‐based Macro-Micro Bridge: $\eta$ Ratio and Capacity Loss}

The two speed-profile scenarios just introduced are evaluated through the \emph{lag-spacing rule} originally proposed by Newell~\cite{newell2002simplified}, which maps a leading‐vehicle trajectory onto the follower’s ``desired’’ path via a fixed time delay~$\tau$ and a spatial shift~$\delta$.  Denoting the leader’s trajectory by $x_{\text{lead}}(t)$, the Newell reference for the follower reads
\begin{linenomath}
  \begin{equation}
    \label{eq:newell_equation}
    x_{\text{Newell}}(t)=x_{\text{lead}}(t-\tau)-\delta ,
  \end{equation}
\end{linenomath}

where the slope $\omega=\delta/\tau$ reproduces the classical triangular \emph{fundamental diagram}. Instantaneous deviations from this reference are captured by the space-time ratio $\eta(t)$-the fraction of the Newell triangle that remains ``filled’’ by the actual follower trajectory. As shown in Fig.~\ref{fig:traj_profile} (a), A, B, C is three corresponding points on ICEV leader, Newell and EV follower. We can get $\eta$ as following equation:

\begin{linenomath}
  \begin{equation}
    \label{eq:eta_def}
    \eta(t)=
    \frac{AC}{AB} = \frac{t_C - t_A}{t_B - t_A} 
       = \frac{t_C - t_A}{\tau}.
  \end{equation}
\end{linenomath}
A value $\eta=1$ implies the follower sits exactly on its Newell line and hence maintains nominal capacity; $\eta<1$ indicates an timid driving style causing extra capacity loss while $\eta>1$ indicates and aggressive driving style.

To translate this microscopic mismatch into a macroscopic performance metric, we integrate the \emph{gap surplus} $\Delta x(t)=x_{\text{Newell}}(t)-x_{\text{follower}}(t)$ over the duration of the manoeuvre~$[t_a,t_b]$:
\begin{linenomath}
  \begin{equation}
    \label{eq:caploss_int}
    \mathcal{L}
      =\int_{0}^{t}\!\!\bigl[x_{\text{Newell}}(t)-x_{\text{follower}}(t)\bigr]\,\mathrm{d}t
      =\int_{0}^{t}\!\!(\eta(t) - 1)\,(\delta+v_0\tau)\,\mathrm{d}t .
  \end{equation}
\end{linenomath}
The resulting quantity $\mathcal{L}$ has units of \emph{vehicle‐seconds} (lane‐metres) and can be interpreted as the \emph{cumulative capacity loss} of one follower during the episode.  Dividing by the manoeuvre length $t_b-t_a$ yields an average flow reduction that can be propagated to the stream level under the assumption of homogeneous follower behaviour.

\subsection{Closed-Form Solutions for the EV-ICEV Car-Following Scenarios}

\begin{figure}[htbp]
  \centering
  \includegraphics[width=0.95\textwidth]{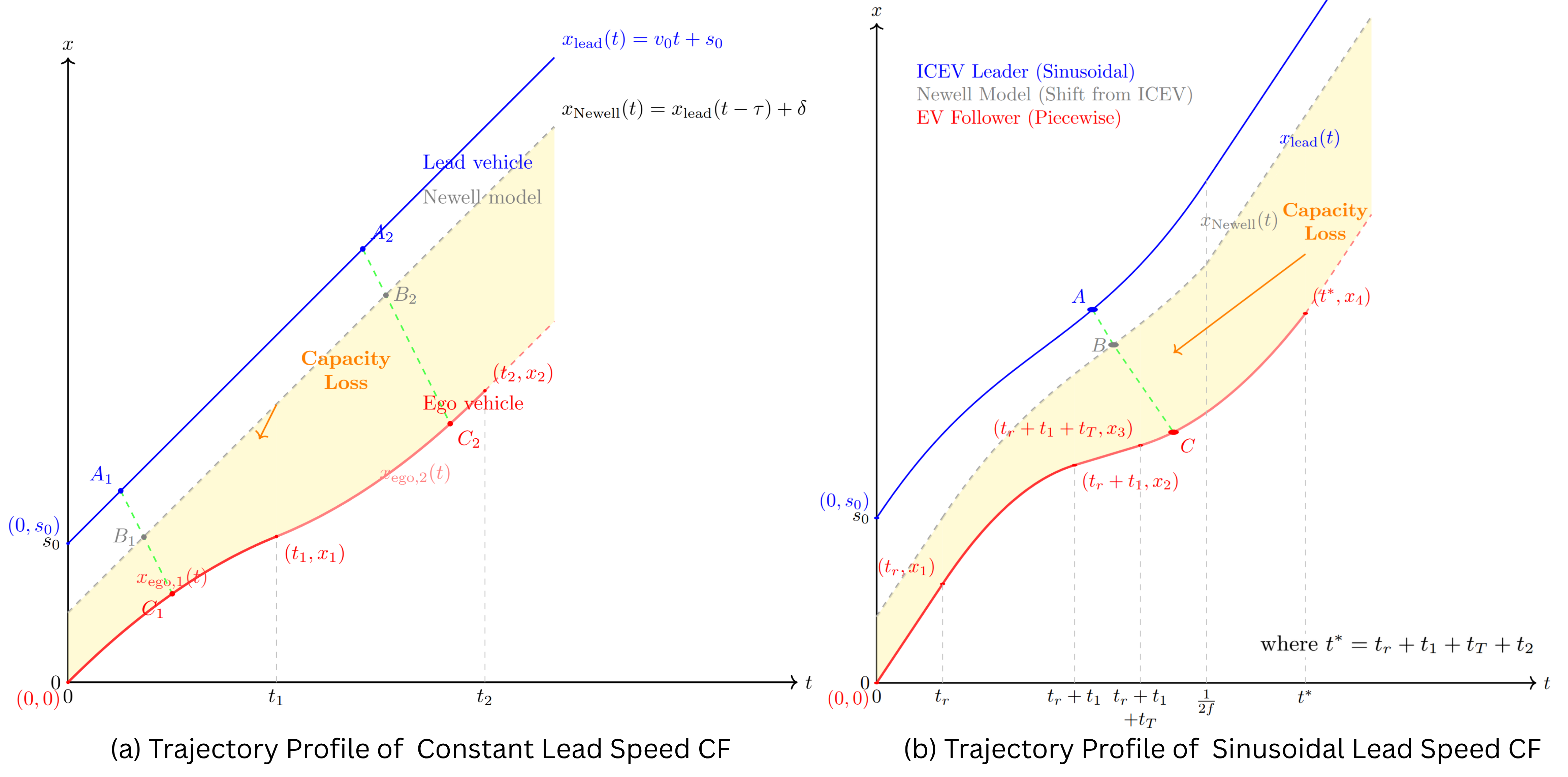} 
  \caption{Trajectory Profile of 2 CF Scenarios; Scenario 1: Constant-speed leader, regen-brake-recover follower; Scenario 2: Leader brake-recover, regen-brake-recover follower. The capacity loss is the yellow area between newell's trajectory and EV follower.}
  \label{fig:traj_profile}
\end{figure}

\noindent
\textbf{Analytic equaction of $\eta$ in Fig.~\ref{fig:traj_profile}(a)}. Given that A and C are any two points on the lead and follower trajectories, respectively, and that the slope of AC is $\omega$, combining Eq.~\ref{eq:traj_leader}, Eq.~\ref{eq:traj_follower}, and Eq.~\ref{eq:eta_def}, we can obtain the expression for $\eta$ for Scenario 1: constant-speed leader, regen-brake-recover follower, as shown in $\eta_1$ in Eq.~\ref{eq:eta_scenario1}.

\begin{linenomath}
  \begin{equation}
    \label{eq:eta_scenario1}
    \eta_1(t)=
    \begin{cases}
      \displaystyle
      \frac{s_0-\tfrac12 a_1\,t^{2}}{\delta+v_0\,\tau},
      & 0 \le t \le t_{1},\\[6pt]
      \displaystyle
      \frac{s_0-\tfrac12 a_1\,t_{1}^{2}-(a_1-a_2)\,t_{1}t-\tfrac12 a_2\,t^{2}}
           {\delta+v_0\,\tau},
      & t_{1} < t < t_{1}+t_{2}.
    \end{cases}
  \end{equation}
\end{linenomath}

\noindent
\textbf{Analytic equaction of $\eta$ in Fig.~\ref{fig:traj_profile}(b)}. Similarly, for any pair of corresponding points A (on the lead trajectory) and C (on the follower trajectory), by combining Eq.~\ref{eq:sinusoidal_lead_traj} and Eq.~\ref{eq:sinusoidal_follower_traj} with the slope $\omega$ of points AC, we can obtain Eq.~\ref{eq:implicit_tA} to describe the corresponding time relationship between points A and C. However, since point A is on a piecewise trajectory, our final $\eta$ analytic expression will also be given in piecewise form. When A is on the sinusodial, it is described by Eq.~\ref{eq:implicit_tA}, and when A is in the uniform velocity segment after recovery, it is described by Eq.~\ref{eq:T_A2}.

\begin{linenomath}
  \begin{equation}
    \label{eq:implicit_tA}
    \Bigl(v_0+\frac{\delta}{\tau}\Bigr)\,t_A
    \;-\;
    \frac{a_0}{2\pi f}\,
    \cos\!\bigl(2\pi f\,t_A\bigr)
    \;=\;
    \frac{\delta}{\tau}\,t_C
    + x_{\mathrm{follower}}(t_C)
    - s_0
    - \frac{a_0}{2\pi f}.
  \end{equation}
\end{linenomath}

where $x_{follower}(t_C)$ is Eq.~\ref{eq:sinusoidal_follower_traj}, which is 5-segment piecewise, and the linear part is shown below:

\begin{linenomath}
\begin{equation}
    \label{eq:T_A2}
    t_A(t) = \frac{x_{follower}(t) + \frac{\delta}{\tau} t - s_0 - \frac{a_0}{\pi f}}{v_0 + \frac{\delta}{\tau}}
\end{equation}
\end{linenomath}
.
\noindent
Because the implicit relation in Eq.~\ref{eq:implicit_tA} contains the transcendental term $\cos(2\pi f\,t_A)$, it possesses \emph{no closed-form algebraic solution} for the intersection time~$t_A$. To obtain an analytic approximation, we linearize~\eqref{eq:sinusoidal_follower_traj} about the zero-order estimate

$
  t_A^{(0)}   =
  \dfrac{\omega t_C+x_{\mathrm{follower}}(t_C)-s_0-\tfrac{a_0}{2\pi f}}
        {v_0+\omega},
$
\noindent
which corresponds to neglecting the cosine term. Applying a single Newton-Raphson iteration yields

\begin{equation}
  t_A^{(1)}
  =
  t_A^{(0)}
  -
  \frac{
        (v_0+\omega)t_A^{(0)}
        -\dfrac{a_0}{2\pi f}\cos\!\bigl(2\pi f t_A^{(0)}\bigr)
        -\omega t_C -x_{\mathrm{follower}}(t_C)+s_0+\dfrac{a_0}{2\pi f}
       }
       {\,v_0+\omega
        +a_0\sin\!\bigl(2\pi f t_A^{(0)}\bigr)} .
  \label{eq:tA_iter}
\end{equation}
\noindent
Hence we adopt the closed-form surrogate

\begin{equation}
  t_A
  \;\approx\;
  \frac{\omega t_C+x_{\mathrm{follower}}(t_C)-s_0-\dfrac{a_0}{2\pi f}}
       {v_0+\omega}
  +
  \frac{\dfrac{a_0}{2\pi f}\,
        \cos\!\Bigl(2\pi f\,
        \dfrac{\omega t_C+x_{\mathrm{follower}}(t_C)-s_0-\tfrac{a_0}{2\pi f}}
              {v_0+\omega}\Bigr)}
       {v_0+\omega}.
  \label{eq:tA_closed}
\end{equation}

\noindent
Substituting~\eqref{eq:tA_closed} into the definition $\displaystyle \eta(t_C)=\dfrac{t_C-t_A}{\tau}$ and recalling
$\omega=\delta/\tau$ gives the approximate \emph{space-time ratio} for the sinusoidal-leader scenario:

%

\begin{linenomath}
  \begin{equation}
    \eta_2(t)\,\approx\,
    \begin{cases}
      \displaystyle
      1-\frac{1}{\tau}
      \Biggl[
        \frac{\omega\,t+x_{\mathrm{follower}}(t)-s_0-\dfrac{a_0}{2\pi f}}
             {v_0+\omega}
        +
        \frac{\dfrac{a_0}{2\pi f}\,
              \cos\!\Bigl(
                2\pi f\,
                \dfrac{\omega\,t+x_{\mathrm{follower}}(t)-s_0-\tfrac{a_0}{2\pi f}}
                      {v_0+\omega}
              \Bigr)}
             {v_0+\omega}
      \Biggr]\\[12pt]
      \displaystyle
        (t-
        \dfrac{%
          x_{\mathrm{follower}}(t)-\omega\,t+s_0+\dfrac{a_0}{\pi f}}%
          {v_0+\omega}) \cdot
      \frac{1}{\tau}
    \end{cases}
    \label{eq:eta_final}
  \end{equation}
\end{linenomath}

\noindent
, where $x_{follower}(t_C)$ is Eq.~\ref{eq:sinusoidal_follower_traj}, which is 5-segment piecewise. The boundary conditions for the simultaneous equations above are $0 \le t_A < \dfrac{1}{2f}$ for the first equation and $t_A \ge \dfrac{1}{2f}$ for the second equation. Due to the complex relationship between $t_A$ and $t$, the boundary conditions for $t$ cannot be given explicitly using analytical expressions. Equation~\eqref{eq:eta_final} is the closed-form expression to evaluate the cumulative capacity loss for the "leader brake-recover, regen brake-recover follower" scenario.

\section{Validation}

In this section, we will use numerical simulation and empirical data analysis to verify the above two situations respectively. Considering that Scenario 1, where the preceding vehicle is traveling at a constant speed, is relatively simple, we will directly use empirical data for verification. For Scenario 2, we will first verify it in simulation, and then find similar scenarios in empirical data for further analysis and verification.

\subsection{Validation of Constant-speed leader, regen-brake-recover follower}

\begin{figure}[h]
  \centering
  \includegraphics[width=0.99\textwidth]{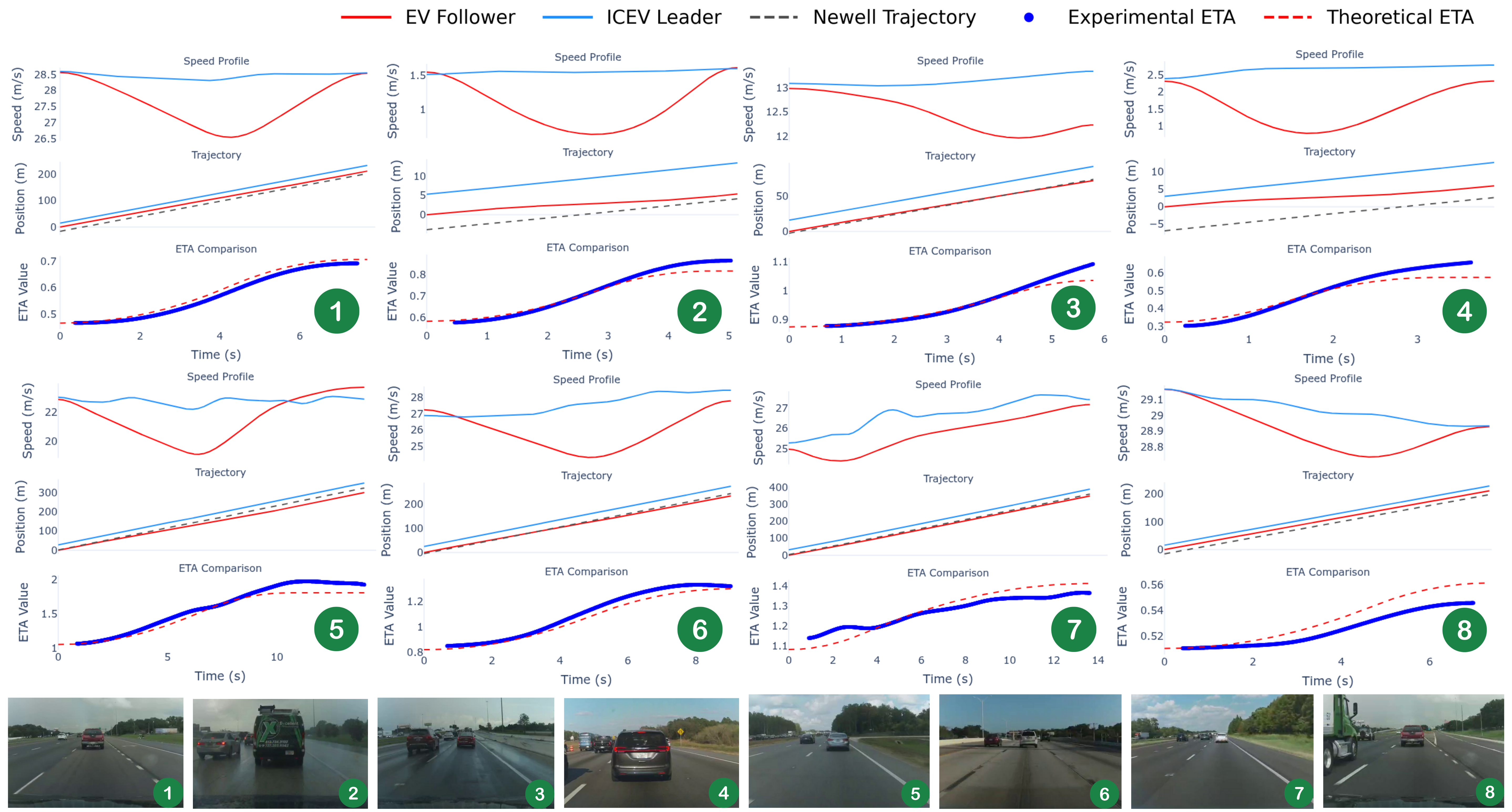} 
  \caption{Empirical Validation for Scenario 1: Constant-speed leader, regen-brake-recover follower; 1, 2, 3, 8 is from KiaNiro; 4, 7 from Hyundai Ioniq-5 One-pedal mode; 5 ,6 from Hyundai Ioniq-5 Strong auto-regen mode.}
  \label{fig:validation1}
\end{figure}

We selected several examples of different electric vehicle models with different regen intensities (Fig.~\ref{fig:validation1}), and intercepted the CF process of the EV first decelerating due to the regen caused by the pedal lift-off and then accelerating back to the element when the leading vehicle is an ICEV and is traveling at an approximately constant speed. We then parameterized the speed profile of the ICEV and the speed profile of the electric vehicle, determined the initial speed and the acceleration at different stages. We then used different methods to calculate the theoretical value and the actual value: For the actual value of $\eta$, we obtained the EV follower trajectory based on the 100Hz EV follower speed integral, and then we obtained the trajectory curve of the leading vehicle based on the 100Hz following distance measured by the sensor. We moved the trajectory curve of the leading vehicle according to the Newell CF Model (Eq.~\ref{eq:newell_equation}) to obtain the Newell curve. Then for each sampling point of 100Hz on the ego vehicle, we took the point through which the slope is calculated the intersection of the line $-\frac{\delta}{\tau}$ with the Newell curve and the ICEV Leader's trajectory curve, thereby obtaining the actual value of $\eta$ according to Eq.~\ref{eq:eta_def}. For the theoretical value of $\eta$, we parameterized the ICEV and EV speed curves and directly substituted them into Eq.~\ref{eq:eta_scenario1}. Fig.~\ref{fig:validation1} clearly shows that although the ICEV Leader and EV Follower in the world only approximate our model definition, the theoretical value of $\eta$ obtained using our method fits the actual value of $\eta$ well. 

\begin{figure}[!h]
  \centering
  \includegraphics[width=0.8\textwidth]{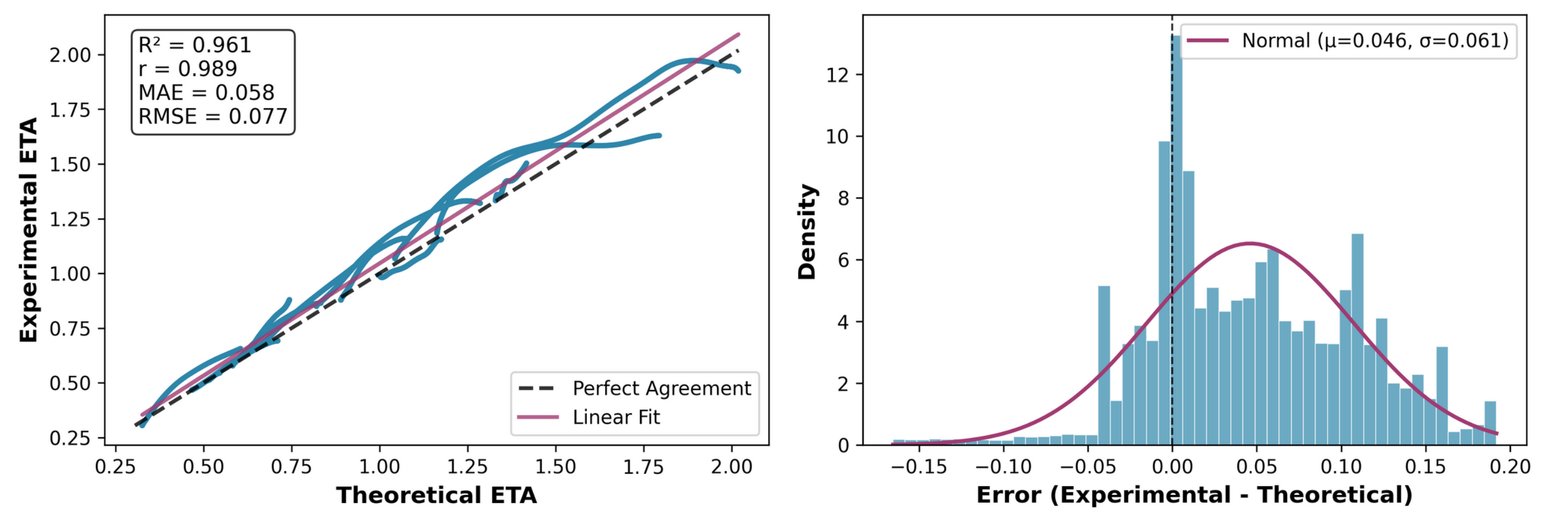} 
  \caption{Validation results for Constant-speed leader, regen-brake-recover follower}
  \label{fig:validation_result_1}
\end{figure}

The results demonstrate that the theoretical model has excellent predictive power, with R² = 0.961, Pearson correlation coefficient r = 0.99, MAE = 0.058, and RMSE = 0.077. The error distribution is approximately normal, with a mean of 0.046 and a standard deviation of 0.061, indicating that the model has an acceptable level of systematic bias. These results fully verify the accuracy and practicality of the established theoretical framework.

\subsection{Validation of Brake-recover leader, regen-brake-recover follower}

We first simulated the EV follower and ICEV leader under different conditions. Considering that the intersection point A of the ICEV Lead in the analytical expression Eq.~\ref{eq:eta_final} may be on the sinusoidal segment or the linear segment, resulting in segmentation, we also use the segmented analytical expression for fitting in this simulation.

\begin{figure}[htbp]
  \centering
  \includegraphics[width=1\textwidth]{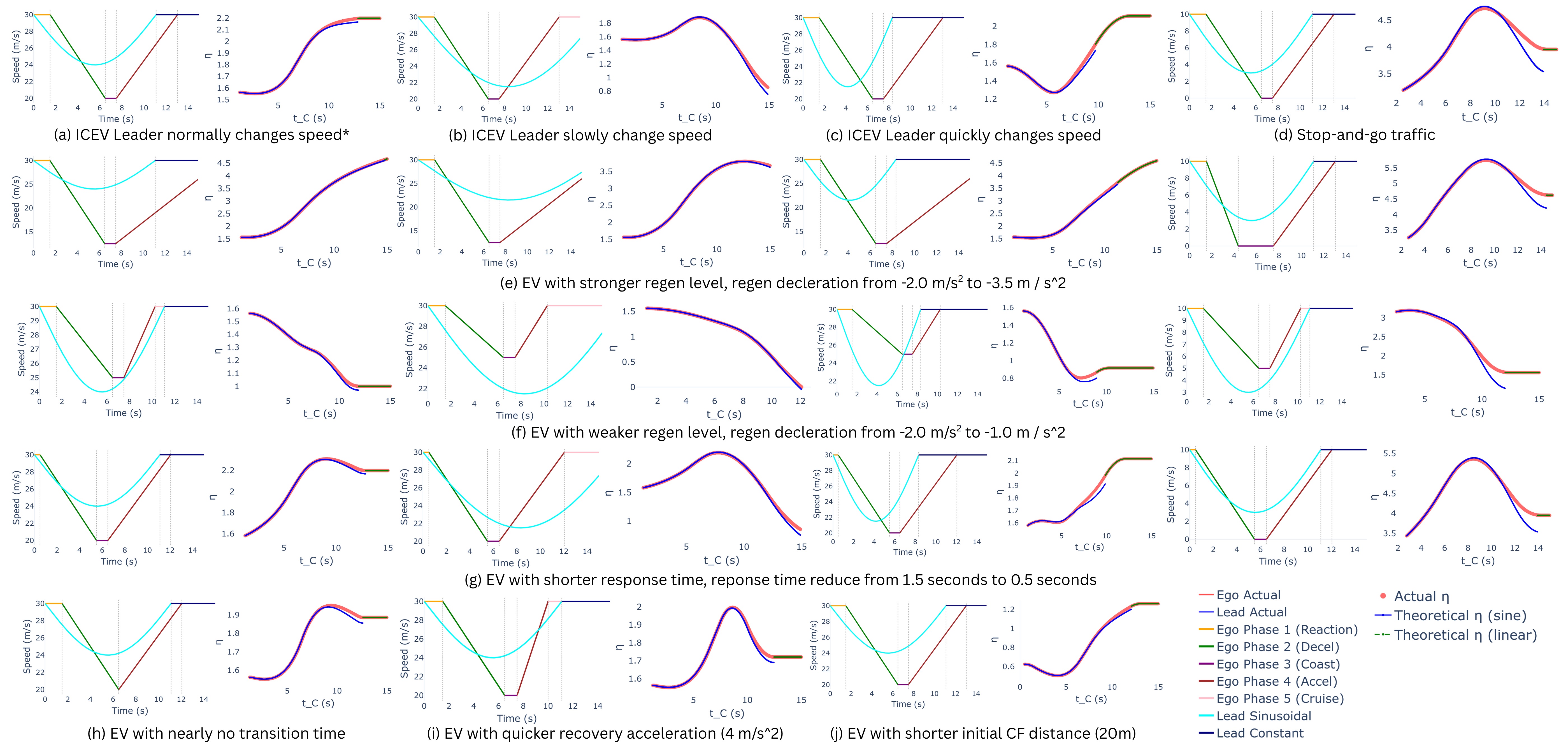} 
  \caption{Comparison of simulation results and actual value of $\eta$ with different initial speed, response time, regen levels, transition time and initial CF distance.}
  \label{fig:validation2}
\end{figure}

As shown in Fig.~\ref{fig:validation2}, we simulated and verified a large number of situation combinations. In Fig.~\ref{fig:validation2} (a), we pre-set $v_0 = 30; a_1 = -2.0; a_2 = 1.8; t_r = 1.5; t_1 = 5; t_T = 1.0; T = 15.0; dt = 0.01; s_0 = 50; a_0 = -6; f = 0.045; \tau=0.8; \delta=8;$ as the baseline case. All values are very common and consistent with our understanding of normal conditions. Then in Fig.~\ref{fig:validation2} (b), (c), and (d), we modified the parameters of the leading vehicle respectively to simulate an ICEV leader (different $f$, $a_0$, $v_0$) that changes speed slowly and a leading vehicle that changes speed quickly. and a low-speed preceding vehicle in stop-and-go scenarios. In Figs.~\ref{fig:validation2} (e), (f), and (g), we simulate the strength of the regen level and the length of the response time required for the driver, operating an EV, to maintain a constant speed after the preceding vehicle decelerates (deceleration from $-2 m/s^2$ to $-3.5 m/s^2$). In Figs.~\ref{fig:validation2} (h), (i), and (j), we simulate the EV's extremely short transition time, faster recovery acceleration, and shorter following distance, respectively. The results show that our equation, Eq.~\ref{eq:eta_final}, can well explain the variations in $\eta$ and provide a good fit to the actual value of $\eta$.

\begin{figure}[htbp]
  \centering
  \includegraphics[width=0.8\textwidth]{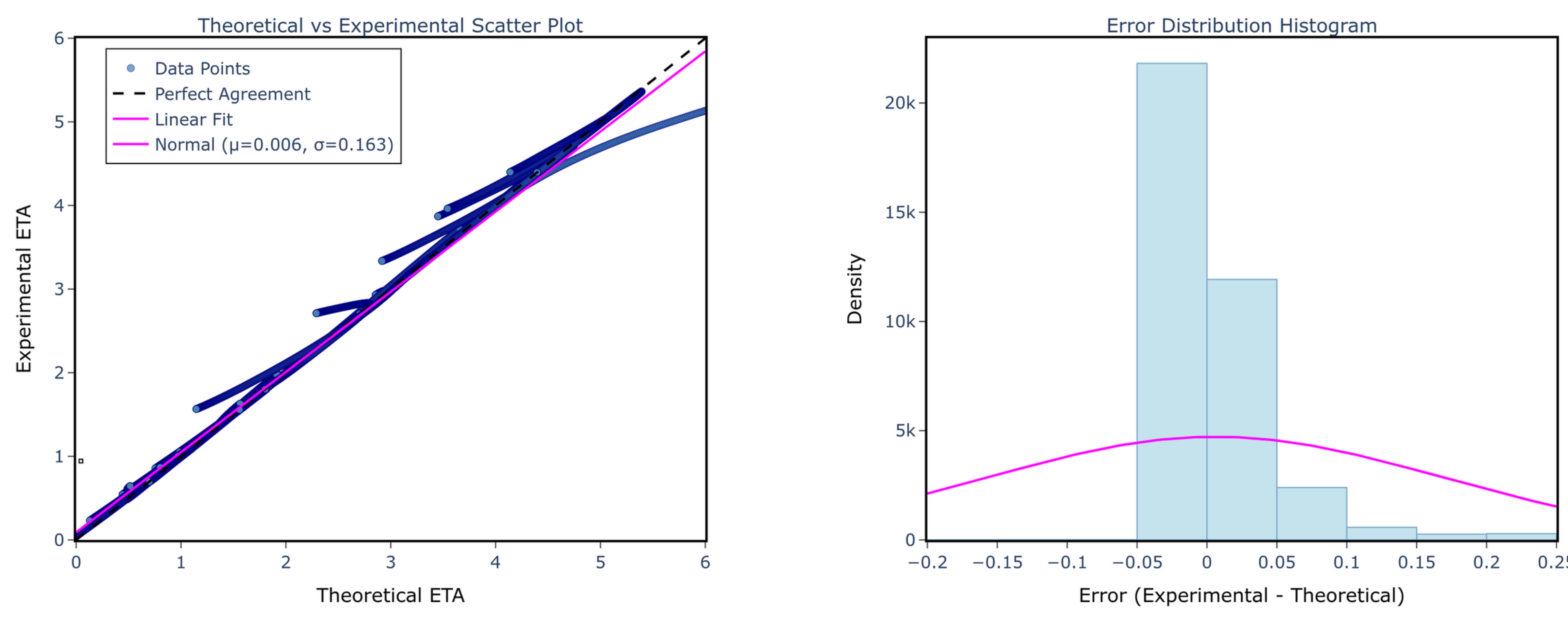} 
  \caption{The simulation results's match with experimental results reaches R²=0.98, showing our analytic equation of $\eta$ fits the Experimental value well.}
  \label{fig:simulation_validation}
\end{figure}

The statistical results in Fig.~\ref{fig:simulation_validation} show a high agreement between the theoretical and experimental values ($R^2$=0.98), with a very small mean error (MAE=0.036). This demonstrates that the theoretical model is highly accurate and reliable, and that our $\eta$ analytical expression can well characterize the true value of $\eta$ in the CF model.

\begin{figure}[htbp]
  \centering
  \includegraphics[width=1.\textwidth]{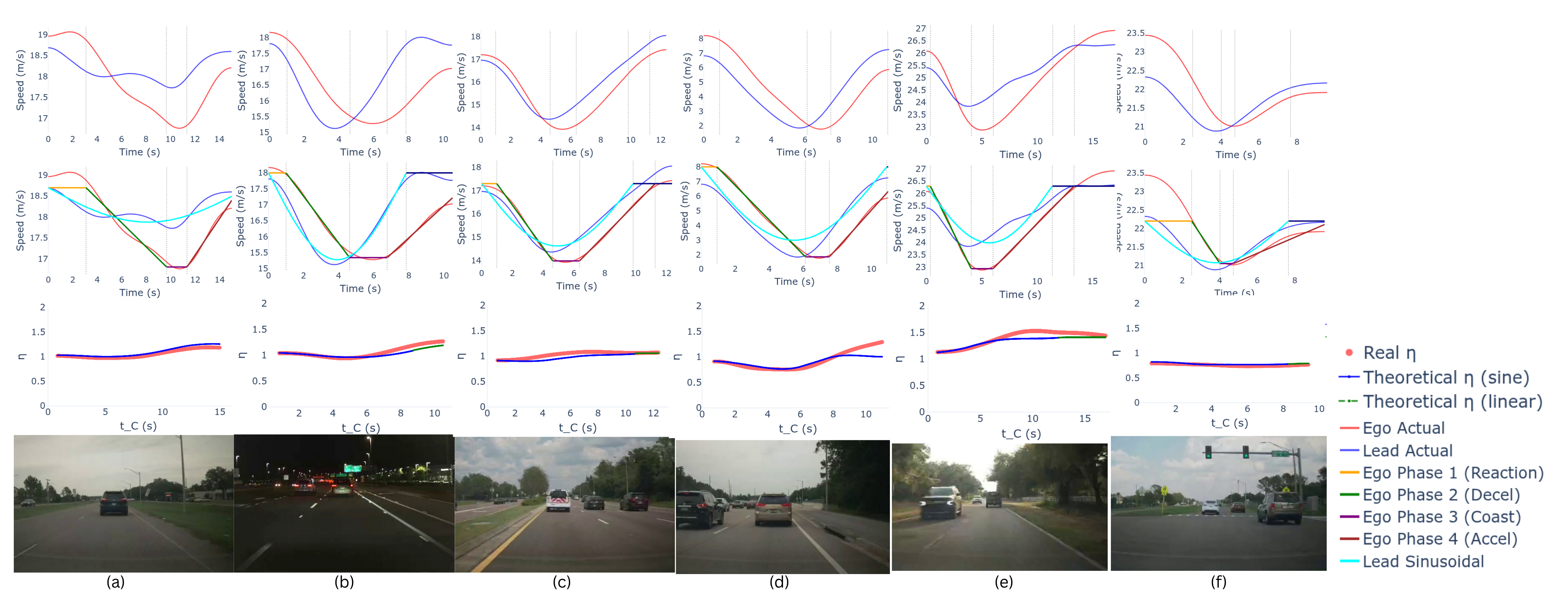} 
  \caption{Empirical Validation from EV dataset. (a) \~ (e) shows a good fit for real world data in EV follows ICEV scenario, and (f) shows the results is also accurate in EV follows EV scenario. }
  \label{fig:empirical_validation}
\end{figure}

Fig.~\ref{fig:empirical_validation} shows several examples from empirical datasets. We first fit the sinusoidal trajectory of the ICEV leader and the piecewise trajectory of the EV follower. We then substitute the fitted parameters into Eq.~\ref{eq:eta_final} to obtain the theoretical value of $\eta$. By comparing the theoretical and actual values, we find that our equation still achieves a good fit.

\section{Discussion}

In this section, we will focus on the relationship between $\eta$ and the different variables describing the physical properties of the term, derived in the previous chapter, to explore how the different characteristics of the term affect traffic flow.

\subsection{The impact of different regen levels on traffic flow}

To explore the effects of different fixed decelerations on $\eta$ caused by different regen levels, we take the partial derivatives of Eq.~\ref{eq:eta_scenario1} and Eq.~\ref{eq:eta_final} with respect to $a_1$. To simplify our results, we let  $\displaystyle  K=v_0+\omega,\;   \Psi(t)=\omega t+x_{\mathrm{follower}}(t)-s_0-\tfrac{a_0}{2\pi f}$.

We finally obtain Eq.~\ref{eq:deta1_da1} and Eq.~\ref{eq:deta2_da1}, corresponding to Case 1: the leading vehicle is traveling at a constant speed; and Case 2: the leading vehicle's sinusoidal velocity curve.

\begin{linenomath}
  \begin{equation}
    \frac{\partial\eta_1}{\partial a_1}(t)=
    \begin{cases}
      -\dfrac{t^{2}}{2(\delta+v_0\tau)}, & 0\le t\le t_1,\\[8pt]
      -\dfrac{\tfrac12\,t_1^{2}+t_1 t}{\delta+v_0\tau},
      & t_1<t<t_1+t_2 .
    \end{cases}
    \label{eq:deta1_da1}
  \end{equation}
\end{linenomath}

\begin{linenomath}
  \begin{equation}
    \frac{\partial\eta_2}{\partial a_1}(t)=
    \begin{cases}
      -\dfrac{1}{\tau K}\,
       \dfrac{\partial x_{\mathrm{follower}}}{\partial a_1}(t)
       \Bigl[1-\dfrac{a_0}{K}\,2\pi f
       \sin\!\bigl(2\pi f\,\Psi(t)/K\bigr)\Bigr],
       & 0\le t_A<\dfrac{1}{2f},\\[12pt]
      -\dfrac{1}{\tau K}\,
       \dfrac{\partial x_{\mathrm{follower}}}{\partial a_1}(t),
       & t_A\ge\dfrac{1}{2f}.
    \end{cases}
    \label{eq:deta2_da1}
  \end{equation}
\end{linenomath}

\noindent
where, $t_A$ is from the definition in Eq.~\ref{eq:eta_final} and the $\frac{\partial x_{\mathrm{follower}}}{\partial a_1}$ is the five-segment follower trajectory derivative  
$\partial x_{\mathrm{follower}}/\partial a_1(t)\ge0$:

\begin{linenomath}
  \begin{equation}
    \frac{\partial x_{\mathrm{follower}}}{\partial a_1}(t)=
    \begin{cases}
      0, & t<t_r,\\[4pt]
      \tfrac12\,(t-t_r)^2, & t_r\le t<t_r+t_1,\\[4pt]
      t_1\,(t-t_r-t_1), & t_r+t_1\le t<t^\ast,\\[4pt]
      t_1\,(t-t^\ast)+\tfrac12\,(t-t^\ast)^2,
        & t^\ast\le t<t^\ast+t_2,\\[4pt]
      t_1 t_2, & t\ge t^\ast+t_2 ,
    \end{cases}
    \qquad t^\ast=t_r+t_1+t_T .
  \end{equation}
\end{linenomath}

The results above (Eq.~\ref{eq:deta1_da1}--\ref{eq:deta2_da1}) shows that \(\partial\eta/\partial\lvert a_1\rvert>0\) throughout every reacting stage of the EV follower.  Because the space–time ratio \(\eta(t)\) measures the excess area between the follower trajectory and its Newell reference, a larger value of \(\eta\) directly translates into enlarged headways and a steeper rise of the cumulative capacity loss integral \(\mathcal{L}\).\footnote{See Eq.~\eqref{eq:caploss_int} for the definition of \(\mathcal{L}\).}  The underlying mechanism is that stronger regenerative braking enforces an 
earlier and more persistent negative acceleration, so the displacement term \(x_{\mathrm{follower}}(t)\) deviates quadratically (during the uniform‐decel 
phase) and then linearly (during the catch-up phase) with respect to \(\lvert a_1\rvert\), a trend consistent with dynamometer evidence on quasi-constant regen deceleration \cite{yang2024regenerative} and with microscopic explanations of stop-and-go growth \cite{laval2010mechanism}.

\begin{figure}[!h]
  \centering
  \includegraphics[width=1.\textwidth]{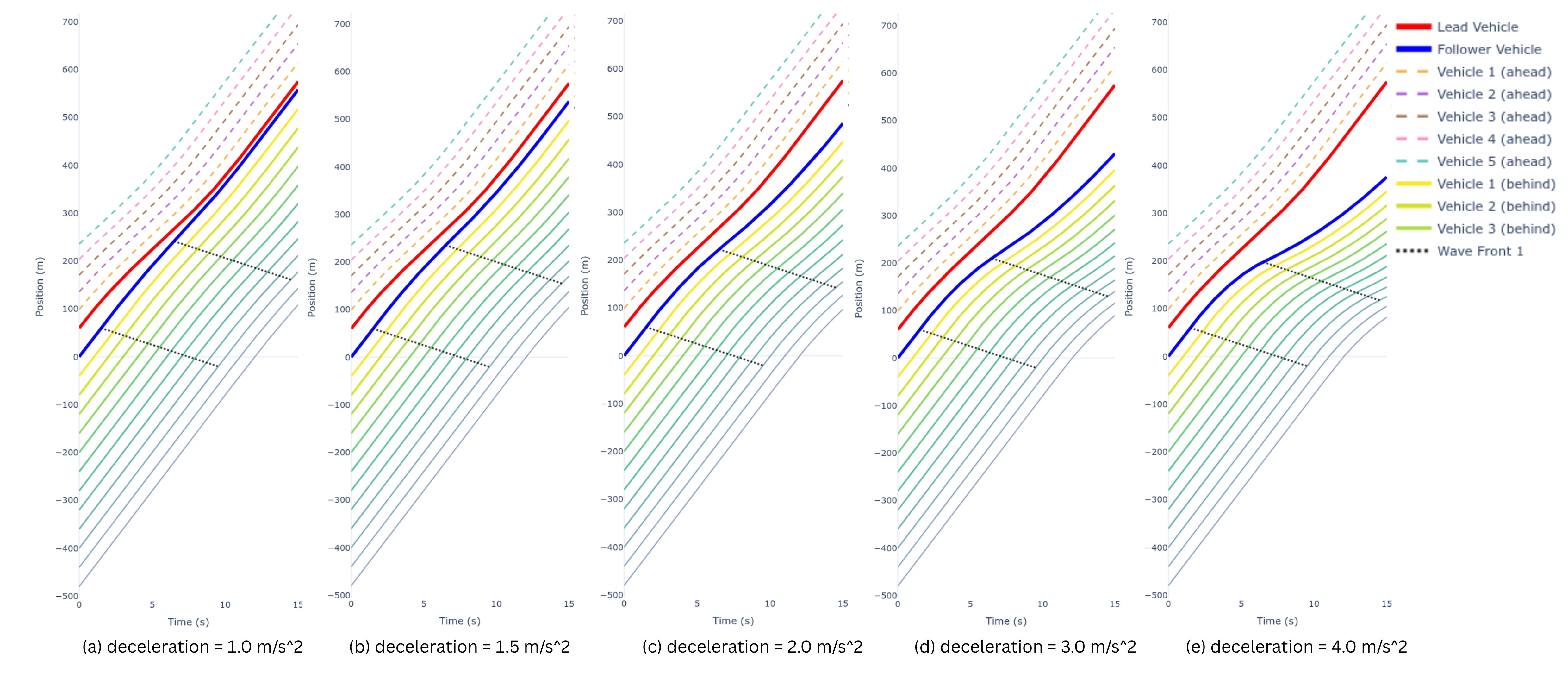} 
  \caption{Simulation for the impact of different regen levels on traffic flow, (a) shows a deceleration at $-1.0m/s^2$, (b) shows a deceleration at $-1.5m/s^2$, (c) shows a deceleration at $-2.0m/s^2$, (d) shows a deceleration at $-3.0m/s^2$, (e) shows a deceleration at $-4.0 m/s^2$.}
  \label{fig:congestion_simulation}
\end{figure}

Increasing the regeneration intensity can recover more energy, but it forces the 
EV to decelerate earlier and harder before congestion, amplifying headway 
fluctuations and reducing effective capacity; therefore, setting a default regen 
gear that is too high in mixed traffic may worsen congestion (Fig.~\ref{fig:congestion_simulation}).

\subsection{The impact of different response time and transition time of EV CF to traffic flow}

We use the same method as when exploring $a_1$ to calculate the partial derivatives of Eq.~\ref{eq:eta_final} with respect to $t_r$ and $t_T$. The results are shown in Eq.~\ref{eq:eta_t_r} and Eq.~\ref{eq:eta_t_T}, respectively.

\begin{linenomath}
\begin{equation}
\label{eq:eta_t_r}
\frac{\partial\eta_2}{\partial t_r}(t)=
-\frac{1}{\tau K}\,
\frac{\partial x_{\mathrm{follower}}}{\partial t_r}(t)
\Bigl[1-\varepsilon(t)\Bigr],\quad
\varepsilon(t)=\frac{a_0}{K}\,2\pi f\,
              \sin\!\Bigl(\tfrac{2\pi f\,\Psi(t)}{K}\Bigr),
\end{equation}
\end{linenomath}

\begin{linenomath}
\begin{equation}
\label{eq:eta_t_T}
\frac{\partial\eta_2}{\partial t_T}(t)=
\begin{cases}
0,& t<t^\ast,\\[4pt]
+\dfrac{v_1}{\tau K}, & t\ge t^\ast,
\end{cases}
\qquad
t^\ast=t_r+t_1+t_T,\quad
v_1=v_0+a_1 t_1 .
\end{equation}
\end{linenomath}

\noindent
,where the piecewise derivative of $x_{follower}$ with respect to $t_r$ is the following equation:

\(\displaystyle
 \frac{\partial x_{\mathrm{follower}}}{\partial t_r}(t)=
 \begin{cases}
   0,&t<t_r,\\[2pt]
   -a_1\,(t-t_r),&t_r\le t<t_r+t_1,\\[2pt]
   -a_1\,t_1,&t_r+t_1\le t<t^\ast,\\[2pt]
   -a_1\,t_1-a_2\,(t-t^\ast),&t^\ast\le t<t^\ast+t_2,\\[2pt]
   -a_1\,t_1-a_2\,t_2,&t\ge t^\ast+t_2 .
 \end{cases}
\)

\noindent
We can further obtain the partial derivatives of capacity loss with respect to $t_T$ and $t_r$ as shown in Eq.~\ref{eq:caploss_t}:

\begin{linenomath}
\begin{equation}
\label{eq:caploss_t}
\frac{\partial\mathcal{L}}{\partial t_r}=
D\!\int_0^{t}\!\frac{\partial\eta_2}{\partial t_r}\,dt' < 0,\qquad
\frac{\partial\mathcal{L}}{\partial t_T}=
D\!\int_0^{t}\!\frac{\partial\eta_2}{\partial t_T}\,dt' > 0,
\quad
D=\delta+v_0\tau .
\end{equation}
\end{linenomath}

In summary, a shorter reaction delay (\(t_r\downarrow\)) increases 
\(\eta\) and thus elevates capacity loss, whereas a shorter coasting period 
(\(t_T\downarrow\)) lowers both \(\eta\) and \(\mathcal{L}\).

\begin{figure}[!h]
  \centering
  \includegraphics[width=1.\textwidth]{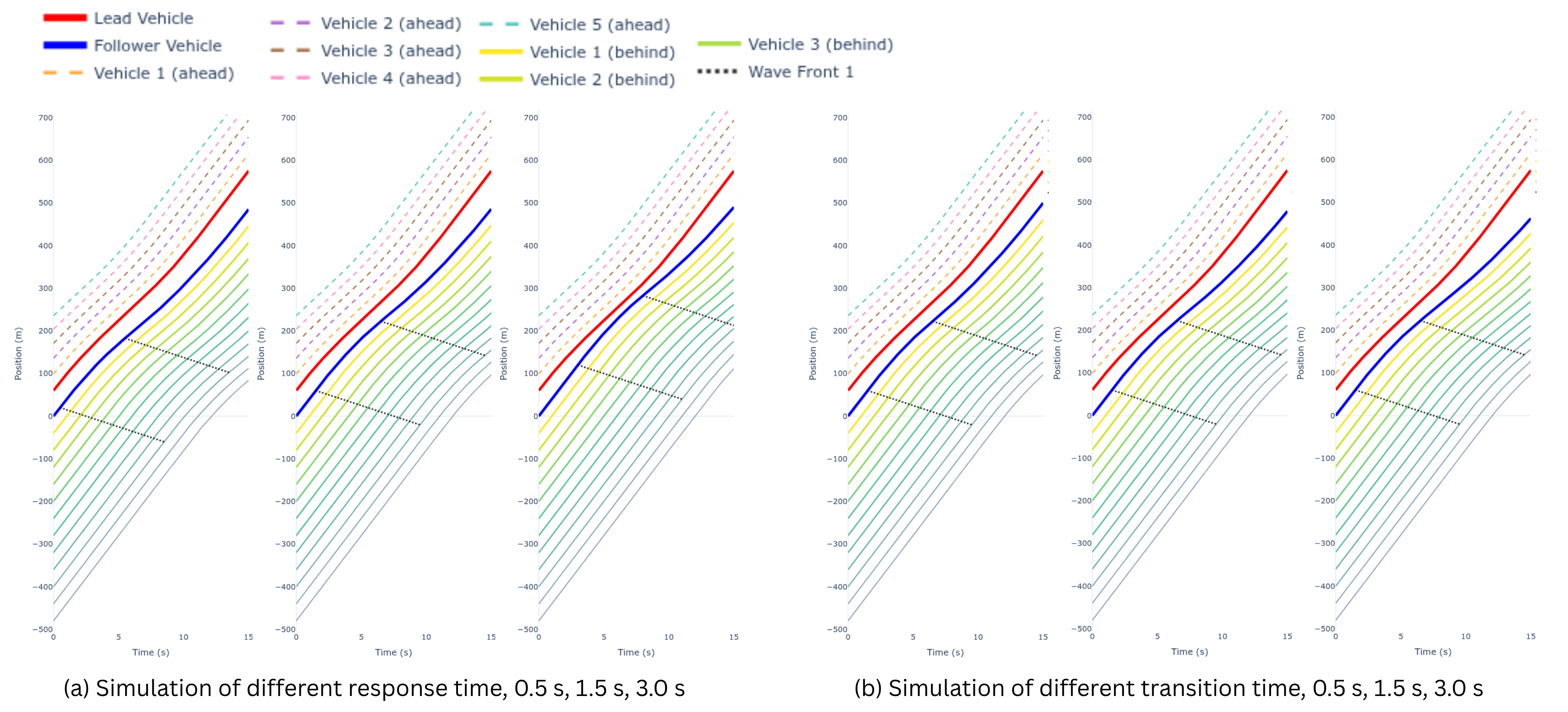} 
  \caption{Simulation for the impact of different response time and different transition time of EV on traffic flow; (a) shows different response time (from left to right) at 0.5, 1.5, 3.0 seconds; (b) shows different transition time (from left to right) at 0.5, 1.5, 3.0 seconds.}
  \label{fig:time_impact}
\end{figure}

The simulation results in Fig.~\ref{fig:time_impact} corroborate the
sensitivity findings of Sect.~\ref{sec:modeling}.  When the reaction delay
\(t_{r}\) is short-\(0.5\;{\rm s}\) in the left‐most panel of
Fig.~\ref{fig:time_impact}\,(a)-the EV follower engages regenerative braking
earlier, lifting the blue trajectory above its Newell reference sooner and
thereby widening the space–time wedge; as \(t_{r}\) is lengthened to
\(3.0\;{\rm s}\), that wedge narrows and both the instantaneous headway and the
cumulative loss shrink.  In contrast, increasing the transition (coasting) time
\(t_{T}\) from \(0.5\) to \(3.0\;{\rm s}\) (Fig.~\ref{fig:time_impact}\,(b))
delays the onset of catch‐up acceleration: the yellow–to–green fan of
follower trajectories shifts right, the vertical separation from the red Newell
line persists longer, and the capacity‐loss polygon enlarges.  Hence, while
aggressive control that minimises \(t_{r}\) enhances responsiveness, it also
magnifies the flow penalty unless the subsequent \(t_{T}\) is kept short; a
balanced tuning of both parameters is therefore essential for trading off
energy recovery, ride comfort, and corridor throughput in mixed EV-ICEV traffic.

\subsection{Conclustion}

This study provides the first physics-grounded link between electric-vehicle
regenerative-braking settings and traffic-flow performance.  Empirical analysis
of 197.5 h of high-resolution trajectories from eight production EV models
confirmed two key kinematic signatures-quasi-constant acceleration/deceleration
and a short, jerk-bounded transition-whose magnitudes scale with user-selected
regen levels.  Embedding these signatures in a Newell-type lag–spacing
framework yielded closed-form expressions for the space–time ratio
\(\eta(t)\) and the associated cumulative capacity loss \(\mathcal{L}\) under
both constant-speed and sinusoidally varying leaders.  Numerical exploration
and dual-source validation (controlled drives and community logs) showed that
stronger regen, longer coasting transitions, and shorter reaction delays all
enlarge \(\eta\) and inflate \(\mathcal{L}\), revealing a tangible trade-off
between energy recovery and corridor throughput.  These insights offer
practical guidance for OEM default settings, adaptive-cruise logic, and
signal-timing policies in mixed EV–ICEV traffic.

\noindent
\textbf{Future work.}  We will continue to refine our instrumentation-adding
rear-facing sensors and synchronised V2X logs-to capture the behaviour of
vehicles \emph{following} an EV and to close the data loop on multi-vehicle
interactions.  The complete, open-source logging suite will be released to a
wider community of volunteers, enabling large-scale, driver-diverse data
collection and deeper investigation of EV penetration effects, multi-lane
interactions, and cooperative control strategies aimed at balancing energy
savings with network capacity.

\newpage

\bibliographystyle{trb}
\bibliography{trb_template}
\end{document}